\documentstyle[preprint,floats,aps,epsf,epsfig,prb]{revtex}
\newlength{\bxwidth}\bxwidth=0.8\textwidth

\newcommand\prm[2]{$ $\vskip 2.4 truein
\centerline{\epsfig{file=#1,width=\bxwidth} }\vskip 0.5truein
\centerline{{\bf Fig.} #2}}
\newcommand\prk[2]{$ $\vskip 2.4 truein\vskip -2 truein
\centerline {\epsfig{file=#1,width=\bxwidth}} \vskip 0.5truein
\centerline{{\bf Fig.} #2}}
\begin{document}
\title{Scaling of the Coercive Field with Thickness in Thin-Film Ferroelectrics}
\author{P. Chandra,$^1$ M. Dawber,$^2$ P.B. Littlewood$^3$ and J.F. Scott$^2$}
\address{$^{1}$Department of Physics, Rutgers University, Piscataway NJ 08855}
\address{$^{2}$Symetrix Centre for Ferroics, Earth Sciences
Department, Downing
Street, University of Cambridge, Cambridge U.K.
CB2 3EQ}
\address{$^{3}$Cavendish Laboratory, University of Cambridge,
Madingley Road, Cambridge U.K.
CB3 OHE}
\maketitle
\vskip 0.35truein
\begin{abstract}
Motivated by the observed thickness-scaling of the coercive field in ferroelectric
films over five decades, we develop a statistical approach towards
understanding the conceptual underpinnings of this behavior.   Here the
scaling exponent is determined by
the field-dependence of a known and measured quantity, the
nucleation rate per unit area.  We end with a discussion of
our initial assumptions and point to instances  where they
could no longer be applicable.
\end{abstract}
\eject
Experimentally the coercive field ($E_c$)
of a thin-film ferroelectric increases  with decreasing sample
thickness ($d$).\cite{Fatuzzo67,Scott00}  Recently we have shown that
$E_c$ displays a specific thickness-dependence, $E_c \sim d^{-2/3}$, over five decades
ranging from 100 microns down to 1 nanometer.\cite{Dawber03}  Here we study the
conceptual underpinnings of this dramatic scaling behavior. Guided by a set of
experimental facts, we develop a statistical approach that yields the expression, $E_c
\sim  d^{-{1\over\alpha}}$.   In particular we find that this exponent $\alpha$ is
determined by the field-dependence of another measured quantity, the nucleation rate
unit area ($N(E)$).  Experimentally it is known\cite{Fatuzzo67,Merz54,Stadler58}
that $\alpha = 3/2$, which then leads to the observed scaling relation, $E_c \sim
d^{-2/3}$.   Finally we emphasize the necessary conditions for this scaling
to occur, indicating situations where it may not be applicable.

The thickness-dependence of the coercive field in ferroelectric films ranging
from 100 microns to 200 nanometers has been known to display the relation
$E_c \sim d^{-2/3}$ for some time.\cite{Scott00} However recent
measurements of $E_c$ for PVDF films down to one nanometer indicate significant
deviation from the expected scaling behavior.\cite{Bune98,Ducharme00}
Elsewhere we have noted that at such small length-scales the imperfect conducting
nature of the capacitor electrodes becomes important.\cite{Dawber03}
More specifically the full charge on each plate does not reside in the
plane of the ferroelectric-electrode interface, completely compensating
for the spontaneous polarization in the ferroelectric film.  Rather it
is distributed over a small but finite region in the electrode, and
there is {\sl incomplete} charge conpensation at the interface.
Physically this results in a depolarization field in the film which
must play a role in the switching process.  Since the potential in the
imperfect but realistic electrodes results from the induced charge density
it will be proportional to the spontaneous polarization $P_s$ in the film.
In many ferroelectric materials $P_s$ is weakly $d$-dependent so that
the depolarization field will scale roughly inversely with $d$; then the associated
corrections will be important in films of decreasing thickness.

The presence of a depolarization potential indicates that the switching field measured
by the external circuit is {\sl not} the same as that in the ferroelectric. More
specifically the coercive field in the film is\cite{Dawber03}
\begin{equation}
E_c = \frac{ V + 8\pi P_s a}{d + \epsilon_f (2a)}
\label{E}
\end{equation}
where $V$ is the voltage drop across the capacitor, $a =
\frac{\lambda}{\epsilon_e}$ is a normalized screening length ($\lambda$) associated with
the charge, and $\epsilon_e$ and $\epsilon_f$
are the dielectric constants of the metal electrodes and the film respectively.
Here we have assumed a Yukawa form for the charge density at the ferrolectric-electrode
interface.  We observe that when $a = 0$ we recover the result that the film and the
measured fields are identical ($E_c = \frac{V}{d}$).  For PVDF the addition of
such depolarization corrections to the measured coercive fields leads to the recovery of
the  scaling behavior. We refer the interested reader elsewhere for more details of
this treatment.\cite{Dawber03}
In Figure 1 we present a normalized scaling plot of
$\log
\frac{E_c}{C}$ vs $\log d$ where $C$
for three distinct material types of ferroelectric films and note that the slope of
this line is $s = -0.66$ for five decades; here $C$ is a
materials-specific thickness-independent normalization constant.  For the PVDF data, we
estimated the screening length associated with the aluminum electrodes to be
$a = 0.45 \AA$ using a Thomas-Fermi
approximation;\cite{Dawber03} in the inset of Figure
1 we display the inverse capacitance vs. thickness data for the PVDF
films,\cite{Bune98,Ducharme00} and note that our fit yields $a = 0.34 \pm 0.15 \AA$ so
that our estimate is within the bounds set by the experiment.

Motivated by the tremendous success of this semi-empirical scaling relation, $E_c \sim
d^{-2/3}$, we now turn towards understanding its conceptual foundations.  First,
we recall that an electric field is proportional to a force and thus is usually an
intensive quantity.  Thus the observation that the coercive field is thickness-dependent
indicates the breakdown of standard self-averaging.  This raises the possibility
that homogenous nucleation might be responsible for the observed $E_c(d)$,
since here rare fluctuations determine physical quantities.
However such processes have been shown to be energetically improbable in
ferroelectrics.\cite{Landauer57} Furthermore we expect that the probability of
having rare fluctuations anywhere in the sample would increase with system size,
resulting in a decreasing $E_c$ with diminishing thickness in
contradiction with observation.

We can posit a voltage drop across a surface layer at the metal-ferroelectric interface;
then the coercive field of the entire ferroelectric film is
thickness-dependent. However the resulting $E_c$ is inversely
proportional to
$d$, which is not in agreement with experiment.\cite{Fatuzzo67}  Noting the observed
scaling behavior,
$E_c \sim d^{-2/3}$, Janovec\cite{Janovec58} and later Kay and Dunn\cite{Kay62} developed
phenomenological appproaches to the problem that involved the  characterization
of rare fluctuations at the metal-electrode boundary. Though conceptually appealing,
their treatments were logically inconsistent since they applied ideas of homogenous
nucleation, with uniform probability of reverse-domain creation throughout the film,
to particular sites on the sample boundary.  The observed
$d$-dependence of $E_c$ is recovered in these papers, and results from
the specific
geometry of the critical nucleus, a half prolate spheroid.\cite{Janovec58,Kay62}
This conceptual pathway to the scaling law is somewhat surprising given the latter's
broad application to materials with rather different microscopic qualities.  Finally it
is important to establish the working limits of this scaling behavior; for example,
clearly it will not be appropriate for bulk crystals.

The coercive field is a kinetic quantity, and the associated switching involves many
complex processes that include the nucleation and growth of a distribution of domains.
Clearly any theoretical attempt to recover the observed scaling behavior must
be guided by experiment in its choice of initial assumptions.  Here we use a statistical
approach to model $E_c$, restricting our attention to its quasi-static amplitude.
Supported by imaging
measurements,\cite{Shur91,Shur96,Ganpule02} we argue
that the
nucleation of ``dissident'' domains in thin ferroelectric films is most probable
at the electrode boundaries.
The reversal of the sample's macroscopic polarization
is initiated by the nucleation and subsequent growth
of many of these independent domains that develop in parallel
(cf. Figure 2). This approach
clearly neglects long-range
interactions in the system, though it can be adapted to include them
at the mean-field level.\cite{Littlewood86,Chandra89}
However we expect that in thin films, the Coulomb force
will be screened by charges
on the ferroelectric-electrode interfaces.
Elastic interactions can have important
scale-dependent effects on the transition
temperatures in these systems,\cite{Zhang01} but
they are not expected to contribute
to $E_c(d)$ since the strain energy is usually weakly dependent
on field direction.

We now show that the observed thickness scaling of the coercive field can be
recovered from an adapted Kolmogorov-Avrami model\cite{Ishibashi95}
of {\sl inhomogenous} nucleation in a confined geometry.
The coercive field is defined by the condition that half
the sample is transformed. More specifically the thickness-dependence
of $E_c$ is
determined by the expression
$U(E,d) \vert_{E=E_c} = \frac{1}{2}$
where $U(E,d)$ is the untransformed sample fraction as a function of
applied (quasi-static) field $E$ and thickness $d$.
In a model of nucleation and growth, the rate of transformation
of a region at time $t$ is governed not only
by the nucleation rate there but also by
the probability that a domain
that was nucleated earlier somewhere else in the sample sweeps into
the region in question.
As mentioned above, there is strong evidence from
experiment\cite{Scott00,Shur91,Shur96} that
nucleation events occur predominantly at the ferroelectric-electrode
interface.  For the large electric fields ($\ge 10$ kV/cm)
corresponding to the thin-films of our study, it is known that
$v_f$, the forward domain velocity, is of the order of the speed
of sound.\cite{Fatuzzo67}  By contrast $v_l$, the lateral domain motion is
field-dependent and is significantly
slower;\cite{Fatuzzo67} furthermore it is not expected
have a strong dependence on film thickness\cite{Fatuzzo67}
and should therefore make a neglible contribution to $E_c (d)$.
In our treatment we will therefore assume that the nucleation
rate, known to increase with increasing field
strength,\cite{Fatuzzo67}
is high enough that the film is transformed by large numbers
of independent nucleation events, rather than by the slow lateral growth
of a few domains (cf. Figure 2); this approximation is
supported by a detailed quantitative analysis\cite{Scott00a} for
PZT films of dimensions comparable to those used in this study.

The rate of change of the untransformed fraction of the sample,
$U(E,d)$, decreases as a function of increasing time, and
is proportional to the
product of
the material available, $U(E,d)$, and the nucleation
rate.
For the homogenous case, where nucleation is
equally probable throughout the sample, this rate is
defined as the nucleation per unit volume multiplied
by a characteristic volume that the critical nucleus has
swept out in a given time.
By contrast for the
interface-specific
inhomogenous nucleation considered here, we are interested
in a nucleation rate per unit area, $N$, and
the lateral area of a domain, $\cal{A}$; experimentally
$N = N(E)$ is known to be field-dependent.\cite{Fatuzzo67}
We therefore obtain a simple expression for the
time-dependence of $U(E,d)$
\begin{equation}
\displaystyle
\frac{dU}{dt} \approx -  U N(E) {\cal {A}}
\label{dU}
\end{equation}
where we implicity assume
that the time associated with the domain to propagate
across the thickness of the film, $t_f$, is shorter than the period
of the a.c. driving electric field.  This time-scale,
$t_f = \frac{d}{v_f}$, is roughly a nanosecond
for a micron-thick film\cite{Scott00} where the forward domain velocity, $v_f$,
is of order the speed of sound and is known to be asymptotically
independent of
field-strength at high fields;\cite{Fatuzzo67,Lee01}
we note that the thickness provides a natural time-scale
that cuts off the time evolution of the system.
The resulting expression for the fraction of the sample
untransformed as function of field and thickness,
can therefore be taken to be quasi-static for all practical
measurements since
domain propagation across the film is effectively
instantaneous \cite{Fatuzzo67,Scott00}.
With the time replaced by the natural time scale $t_f$, we have
\begin{equation}
U(E,d) \sim \exp - N(E) \left( \frac{{\cal{A}}d}{v_f} \right).
\label{U}
\end{equation}
It is important to note that within this framework
many independent nucleation and growth events occur in
parallel, so that the detailed nature of the final state is not
crucial (cf. Figure 2).
In Figure 2 we assume columnar grains
for graphical simplicity;\cite{Duiker90}
there the shaded
regions refer to the untranformed parts of the sample.

The equation (\ref{U}) is valid with the implicit underlying assumption
that nucleation dominates over lateral growth.  In order to
obtain a more quantitative condition for  the validity of this expression,
we can do a direct comparision of these two different rates.
More specifically, the change in time of the transformed fraction by
lateral growth is proportional to the number of transformed domains
$(\frac{1-U}{\cal{A}})$ and their lateral velocity ($\frac{d{\cal A}}{dt}$)
which results in
\begin{equation}
\frac{d(1 - U)}{dt} = (1 - U) \frac{d \ln \cal{A}}{dt}
\label{growth}.
\end{equation}
An expression for a similar change due to nucleation is
simply proportional to the untransformed fraction and the nucleation
rate and is
\begin{equation}
\frac{d(1 - U)}{dt} = U N {\cal {A}}.
\label{x}
\end{equation}
Then in order for lateral growth to be less important than nucleation
we must have
\begin{equation}
\frac{d \ln {\cal A}}{dt} << N {\cal {A}}
\label{condition}
\end{equation}
where we take $U \sim \frac{1}{2}$ since this is its value
when $E \sim E_C$.  Therefore (\ref{condition}) is our working premise
for the validity of equation (\ref{U}) and all that follows from it.
We note that recent atomic force microscopy studies of
switching in thin ferroelectric
films support our contention that it is dominated by domain forward growth
processes.\cite{Molotskii03}

For field-strengths greater than $10 kV/cm$, relevant for the thin-films
studied here,
the nucleation
rate per unit area is known to have a power-law dependence
on applied field.\cite{Fatuzzo67}  If we accordingly
put $N(E) \sim E^\alpha$ in (\ref{U}),
the expression for the thickness-dependence of the coercive field,
$U(E_c,d) = \frac{1}{2}$,
becomes $E_c^\alpha d \sim 1$; this then yields
\begin{equation}
E_c \sim
d^{-\frac{1}{\alpha}}.
\end{equation}
Experimentally\cite{Merz54} and theoretically,\cite{Stadler58}
it is known from switching kinetics
that $\alpha = \frac{3}{2}$ over a broad range of fields;
this value of $\alpha$ then
leads to the result $E_c \sim d^{-2/3}$.
We note that the scaling of $E_c$ with thickness
follows directly from the field-dependence of the density
of nucleation sites, $N(E)$, which is
a {\sl statistical} quantity that could plausibly  have similar behavior
for materials with different microscopic features.

We have therefore successfully recovered the observed thickness scaling
of the coercive field using a statistical approach that is not dependent
on particular sample specifics.  It is important to be clear about
our underlying assumptions, and to point to situations where
they could break down.  First of all, we have assumed that nucleation
of new reverse-polarization domains occur solely at the ferroelectric-electrode
interface.  Next we have considered situations where the transformation of
the new polarization domains occurs predominantly by nucleation and {\sl not}
by lateral growth of preexisting ``dissident'' ones.  Finally we assume that,
once formed, the new domains grow unimpeded to the other side of the film.
Clearly these assumptions will not be valid for all ferroelectric capacitors.
For example it is known\cite{Scott00a} that for thicker samples than those considered
here, the switching process is dominated by lateral growth. The specific
length-scale where this occurs depends on material-specific
quantities.\cite{Scott00a}
Similarly in disordered and granular films, grain boundaries and charge vacancies could
act as nucleation centers away from the sample edges.\cite{Ganpule02}
Finally defects in the ferroelectric could obstruct the motion of
domains, and then again  we would not expect the scaling relation
discussed here to be appropriate.

In summary, we have presented a statistical model that leads to a
specific thickness-dependence of the coercive field.  This scaling behavior
is observed over five decades of length-scale in several different
materials.\cite{Dawber03}  Guided by a number of experimental observations,
we recover the observed scaling by assuming that nucleation at the electrode
interface dominates the switching process, and that the resulting ``dissident''
domains propagate uninhibited across the film's thickness.  The exponent
associated with the functional form of $E_c(d)$ is related to another known and
measured exponent, that associated with the field-dependence of
the nucleation rate per unit area.  We emphasize that $N(E)$ is a quantity that is not
dependent on sample microscopics, and thus our result is  compatible with experimental
findings.  We note that there remain a number of unresolved theoretical issues
associated with this derivation of $E_c$ that include the calculation of
realistic energy barriers and the determination of the absolute rather than the relative
magnitude of $E_c$.  Finally we would like to study the behavior of
the coercive field at finite-frequency, and in this case it may not be possible to
neglect the long-range elastic and Coulomb interactions as we have done here.

We thank K. Rabe for discussions.


\newpage

\noindent{\bf Figure Captions}

\renewcommand{\labelenumi}{{\bf Fig.} \theenumi .}
\begin{enumerate}

\item
{\bf (Main Figure)} The normalized corrected log coercive field vs. log
thickness which indicates scaling over five decades of
length-scale.  The slope is
$-0.66\pm0.02$, in agreement with the theoretical value of
$-\frac{2}{3}$. {\bf (Inset)} Inverse-capacitance vs.
thickness of PVDF data\cite{Bune98} where the y-intercept
is related to the screening length used in the depolarization
correction.

\item
A schematic of the statistical approach to the scaling
law.  The sample is originally poled (a). Application
of a small electric field leads to a number of nucleation
events at the electrodes, $N(E)$, which then grow across
the same.  For $E < E_c$, the fraction of the sample
that is untransformed (shaded regions), $U(E)$, is more than one-half.
The nucleation density of reverse-polarization domains
increases with increasing applied field, $N(E) \sim E^\alpha$.
For $E = E_c$, one-half of the system remains untransformed.
Details of the domain structure of the final state
are not important, and the thickness-dependence of $E_c$
is determined by the field-dependence of the nucleation
density.

\end{enumerate}

\newpage

\prk{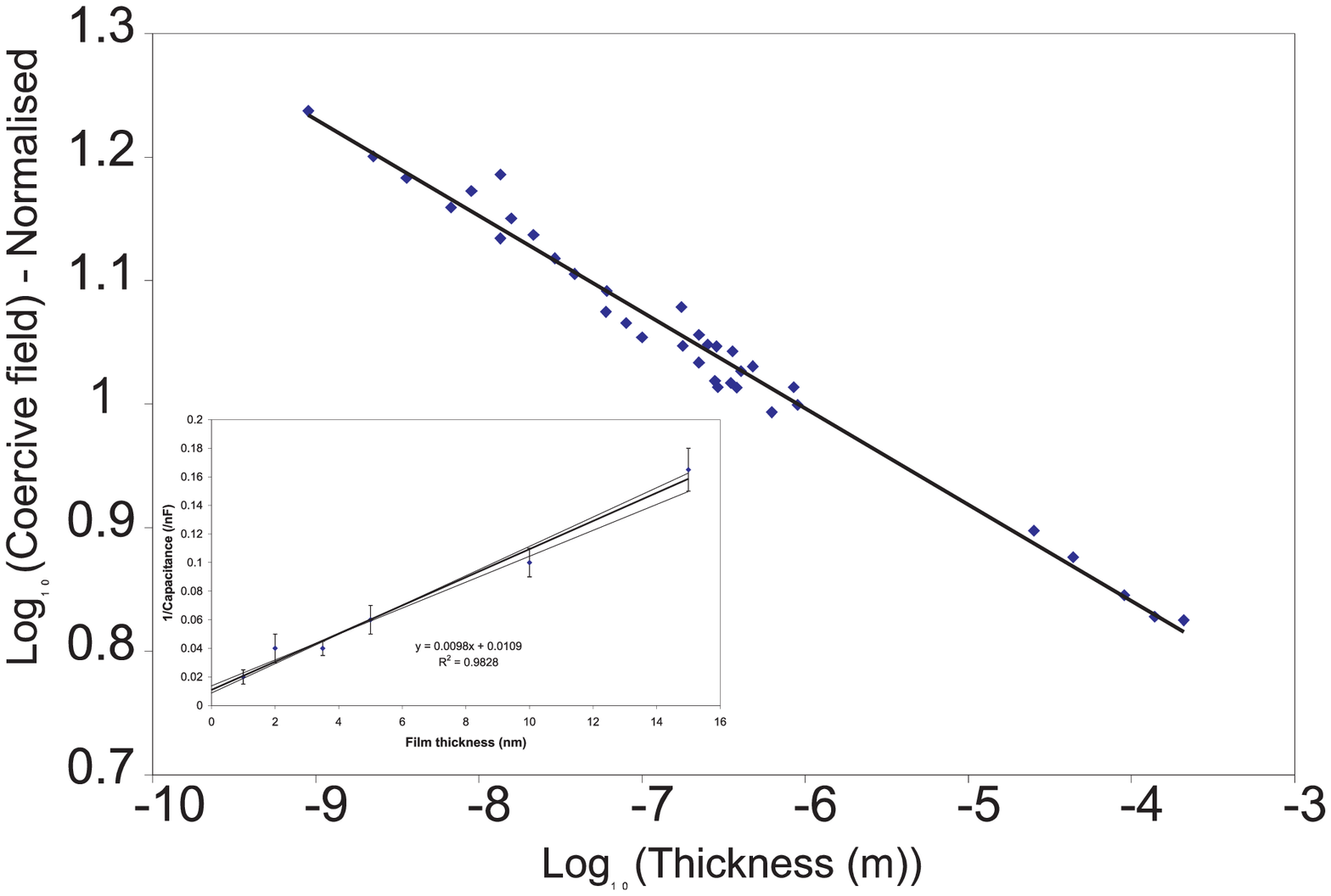}{1}

\prm{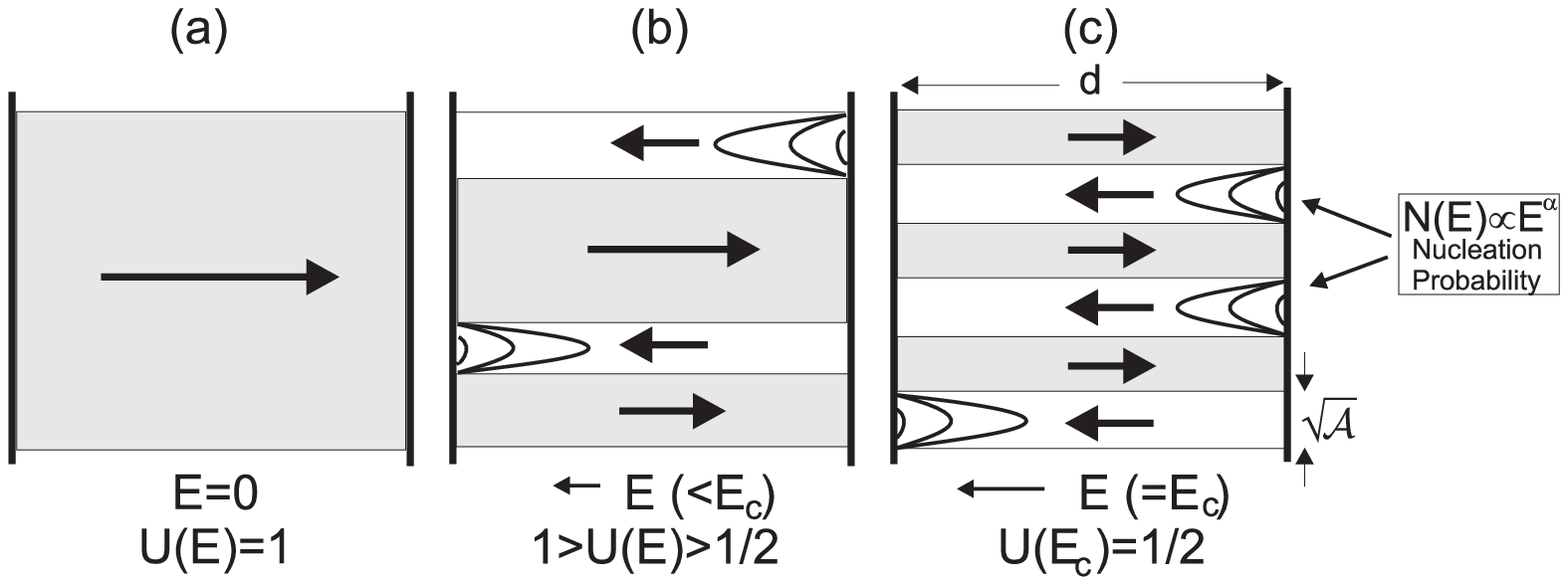}{2}

\end{document}